\documentclass[journal,10pt,letterpaper]{IEEEtran}
\usepackage[utf8x]{inputenc}
\usepackage[T1]{fontenc}
\usepackage{graphicx}
\usepackage{subfigure}
\usepackage{amsmath}
\usepackage{color}
\usepackage{mathptmx}
\usepackage{textcomp}

\begin{document}

\title{Frame Coalescing in Dual-Mode EEE}

\author{Sergio~Herrería-Alonso, 
  Miguel~Rodríguez-Pérez,~\IEEEmembership{Member,~IEEE,} \\
  Manuel~Fernández-Veiga,~\IEEEmembership{Senior Member,~IEEE,} 
  and~Cándido~López-García}


\maketitle

\begin{abstract}
  The IEEE has recently released the 802.3bj standard that defines two
  different low power operating modes for high speed Energy Efficient
  Ethernet physical interfaces (PHYs) working at $40$ and
  $100\,$Gb/s. In this paper, we propose the use of the well-known
  frame coalescing algorithm to manage them and provide an analytical
  model to evaluate the influence of coalescing parameters and PHY
  characteristics on their power consumption.
\end{abstract}

\begin{IEEEkeywords}
  Energy efficiency, IEEE 802.3bj, Energy Efficient Ethernet.
\end{IEEEkeywords}

\section{Introduction}
\label{sec:intro}

To reduce energy consumption of Ethernet links, the IEEE published in
2010 the IEEE~802.3az standard~\cite{802.3az}, known as \emph{Energy
  Efficient Ethernet} (EEE). This norm provides a new operating mode
to be used in Ethernet physical interfaces (PHYs) when there is no
data to transmit. When PHYs are in this low power idle (LPI) mode,
they only draw a small fraction of the power needed for normal
operation, although they are unable to send traffic through their
attached links. Probably, the most natural way to manage
EEE~interfaces consists of entering~LPI whenever the transmission
buffer becomes empty and restoring normal operation when there is new
traffic to transmit. However, this approach is not very efficient
since PHYs consume about the same power during state transitions
(to/from the LPI mode) as in the active state and transition times are
of the same order than a single frame transmission time. In fact,
energy savings can be greatly improved if the number of state
transitions is significantly reduced, for example, just making that
PHYs wait to first accommodate a few frames in the transmission buffer
before exiting LPI (\emph{frame coalescing}). EEE~has shown to be very
effective to reduce energy consumption of $100\,$Mb/s, $1\,$Gb/s and
$10\,$Gb/s Ethernet links, specially when some coalescing control
policy is
applied~\cite{christensen10:_the_road_to_eee,herreria12:gig1}.

The problem of relatively long transition times is even more severe in
$40\,$Gb/s and $100\,$Gb/s Ethernet PHYs since, under these higher
rates, transmission times are significantly lower while transition
times remain similar. This issue has been recently addressed in the
IEEE 802.3bj amendment~\cite{802.3bj} that defines two different low
power modes for high speed interfaces: \emph{Fast-Wake} and
\emph{Deep-Sleep}. In the Fast-Wake state, only some PHY components
can be turned off since clock synchronization must be maintained for
keeping attached links aligned. As a result, this mode requires very
short transition times to resume normal operation but it still draws a
significant portion of the power consumed when active
(70-80\%)~\cite{barrass12:_options_100g}. On the contrary, PHYs in the
Deep-Sleep state consume a very little amount of energy since all
signaling is stopped (10-20\%) but this mode requires considerably
longer transition times.

\cite{mostowfi15:dual_mode_eee}~shows that combining these two low
power modes in high speed EEE links allows energy savings that may be
not achievable with just a single LPI mode. However, this simulation
study assumes that PHYs in any of the two low power modes turn back to
active as soon as new traffic is ready for transmission. In this
paper, we develop an analytical model to evaluate the energy savings
that can be obtained when dual-mode PHYs apply frame coalescing and
wait to queue some frames in the transmission buffer before returning
to active.

The rest of the paper is organized as
follows. Section~\ref{sec:coalescing} presents the basic operations of
dual-mode EEE interfaces using frame coalescing. In
Section~\ref{sec:energy_model} we develop an analytical model to
compute their power consumption. We next particularize this model to
Poisson traffic in Section~\ref{sec:poisson_model}. In
Section~\ref{sec:evaluation}, we validate our analysis through
simulation. Finally, the conclusions are summarized in
Section~\ref{sec:conclusions}.

\section{Dual-Mode EEE Operations}
\label{sec:coalescing}

\begin{figure*}[t]
  \centering
  \resizebox{0.85\textwidth}{!}{\input{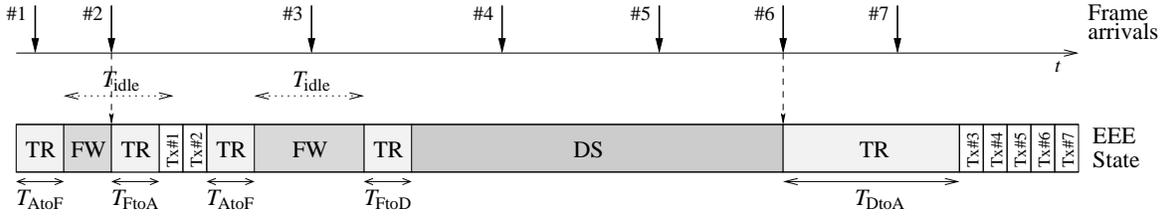}}
  \caption{Dual-mode EEE operations. Example with $Q_{\mathrm{f}}=2\,$frames and $Q_{\mathrm{d}}=4\,$frames.}
  \label{fig:coalescing}
\end{figure*}

Figure~\ref{fig:coalescing} depicts an example of the main operations
of dual-mode EEE~interfaces. Clearly, for maximizing energy savings, a
dual-mode EEE interface should be put to sleep every time its
transmission buffer gets empty. Then, after a short transition of
length~$T_{\mathrm{AtoF}}$, the interface enters the Fast-Wake
mode. \cite{mostowfi15:dual_mode_eee} assumes that the interface will
remain in Fast-Wake until a frame arrives or for a maximum period of
length~$T_{\mathrm{idle}}$. In the former case, the interface would
directly return to the active state after a transition of
length~$T_{\mathrm{FtoA}}$ while, in the latter one, it would
transition to Deep-Sleep after a $T_{\mathrm{FtoD}}$~period. As
suggested in the Introduction, frame coalescing can be applied to
these interfaces to improve their energy savings. Consequently, we
propose that PHYs switch to Deep-Sleep as long as less than
$Q_{\mathrm{f}}$~frames arrive during the
$T_{\mathrm{idle}}$~period. Otherwise, they would resume normal
operation when the $Q_{\mathrm{f}}$-th~frame arrives.

On the other hand, \cite{mostowfi15:dual_mode_eee} also assumes that
the interface abandons the Deep-Sleep mode as soon as a new frame
arrives. Again, we can apply frame coalescing to this mode so that the
interface remain in Deep-Sleep until $Q_{\mathrm{d}}$~frames are
buffered for transmission. We assume that $Q_{\mathrm{f}} \le
Q_{\mathrm{d}}$ to avoid useless transitions to
Deep-Sleep. Eventually, a long transition of
length~$T_{\mathrm{DtoA}}$ will be required to return to the active
state.


Clearly, coalescing frames into bursts increases their delay. So, to
avoid excessive delays, the maximum time an interface can be in a low
power mode (since the first frame is buffered for transmission) should
be limited.

\section{Energy Consumption Model}
\label{sec:energy_model}

In this section we will develop an analytical model to evaluate the
influence of coalescing parameters and PHY~characteristics on energy
consumption. We assume that frame arrivals follow a general
distribution with independent and identically distributed interarrival
times $I_n$, $n=1,2,\ldots$, and average arrival rate~$\lambda$. We
also assume that service times follow an arbitrary distribution
function with mean service rate~$\mu$. Obviously, the utilization
factor $\rho=\lambda/\mu$ must be less than~$1$ to assure system
stability. Finally, we assume that the interface has a transmission
buffer of infinite capacity.\footnote{Note that, if the queue
  thresholds are chosen carefully, additional frame losses should be
  negligible. For example, an interface transitioning to awake from
  Deep-Sleep will only receive, in average, $\lambda
  T_{\mathrm{DtoA}}$~more frames before it is active again. Therefore,
  the buffer should be correctly dimensioned to accommodate, at least,
  $Q_{\mathrm{d}} + \mu T_{\mathrm{DtoA}}$ frames plus a safety
  margin.}

\begin{figure*}[t]
  \centering
  \resizebox{0.8\textwidth}{!}{\input{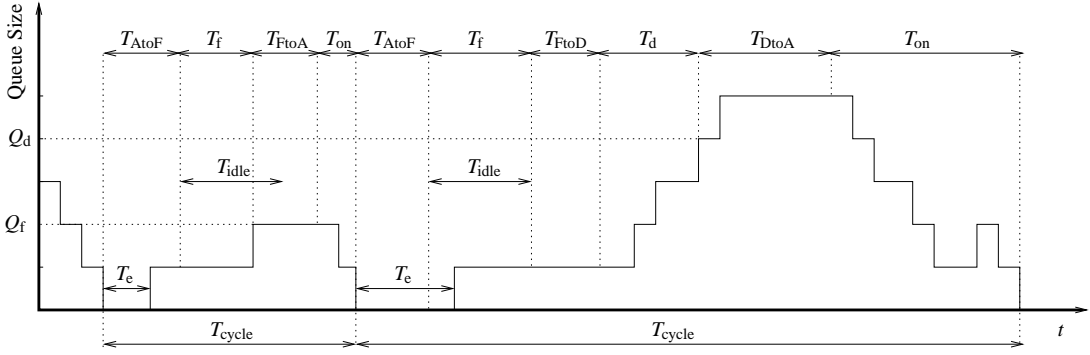}}
  \caption{Coalescing cycles with $Q_{\mathrm{f}}=2\,$frames and $Q_{\mathrm{d}}=4\,$frames.}
  \label{fig:cycle}
\end{figure*}

Fig.~\ref{fig:cycle} shows how the transmission buffer of a dual-mode
EEE~interface evolves. Time intervals when frames are not being
transmitted are \emph{inactive periods} and may be composed of several
\emph{transition periods} ($T_{\mathrm{AtoF}}$, $T_{\mathrm{FtoA}}$,
$T_{\mathrm{FtoD}}$, $T_{\mathrm{DtoA}}$) and two \emph{sleeping
  periods} ($T_{\mathrm{f}}$, $T_{\mathrm{d}}$) in which the interface
is in one of the two defined low power modes. Within the inactive
period there is also an \emph{empty period} $T_{\mathrm{e}}$ that
comprises the time elapsed between the interface is put to sleep and
the arrival of the subsequent first frame. During this empty period no
frames are buffered in the transmission queue of the interface. The
time interval when the interface is transmitting frames is the
\emph{busy period} $T_{\mathrm{on}}$. Finally, an inactive period
followed by a busy period forms a \emph{coalescing cycle}
$T_{\mathrm{cycle}}$. 


\subsection{Energy Consumption}
\label{sec:energy}

Let $\mathrm{E}[P]$ be the mean power consumption for a given
interface that never enters a low power mode and
$\mathrm{E}[P_{\mathrm{EEE}}]$ the mean power consumption for a
dual-mode EEE~interface. Clearly, $\mathrm{E}[P_{\mathrm{EEE}}]$
depends on the proportion of time the interface spends in each
possible state:
\begin{IEEEeqnarray}{rCl}
  \label{eq:P_EEE} 
  \mathrm{E}[P_{\mathrm{EEE}}] &=& \rho_{\mathrm{f}}
  \mathrm{E}[P_{\mathrm{f}}] + \rho_{\mathrm{d}}
  \mathrm{E}[P_{\mathrm{d}}] + \rho_{\mathrm{tr}} \mathrm{E}[P] +
  \rho_{\mathrm{on}} \mathrm{E}[P] \nonumber \\ 
  &=& \rho_{\mathrm{f}}  \mathrm{E}[P_{\mathrm{f}}] + 
  \rho_{\mathrm{d}} \mathrm{E}[P_{\mathrm{d}}] + 
  (1 - \rho_{\mathrm{f}} - \rho_{\mathrm{d}})  \mathrm{E}[P],
\end{IEEEeqnarray}
where $\mathrm{E}[P_{\mathrm{f}}]$ and $\mathrm{E}[P_{\mathrm{d}}]$
are the mean power consumed in the Fast-Wake and the Deep-Sleep modes,
and $\rho_{\mathrm{f}}$, $\rho_{\mathrm{d}}$, $\rho_{\mathrm{tr}}$ and
$\rho_{\mathrm{on}}$ are, respectively, the fractions of time in which
the interface is in Fast-Wake, Deep-Sleep, transitioning between
states and awake. Note that it is assumed that the interface consumes
about the same power during transitions as in the active state since
many components of the transceivers have to be operative during the
state changes. Immediately, the energy consumed on a dual-mode
EEE~interface compared with that consumed on an interface that it is
always active is given by
\begin{equation}
  \label{eq:varphi}
  \varphi = \frac{\mathrm{E}[P_{\mathrm{EEE}}]}{\mathrm{E}[P]} = 1 -
  (1 - \varphi_{\mathrm{f}}) \rho_{\mathrm{f}} - (1 -
  \varphi_{\mathrm{d}}) \rho_{\mathrm{d}},
\end{equation}
where $\varphi_{\mathrm{f}} =
\mathrm{E}[P_{\mathrm{f}}]/\mathrm{E}[P]$ and $\varphi_{\mathrm{d}} =
\mathrm{E}[P_{\mathrm{d}}]/\mathrm{E}[P]$ are the portions of the
active mode energy consumption demanded when the interface is in
Fast-Wake and Deep-Sleep, respectively, and shape the efficiency
profile of the interface.

The factor $\rho_{\mathrm{f}}$ can be calculated as the ratio between
the mean duration of a Fast-Sleep period and the mean duration of a
coalescing cycle:
\begin{equation}
  \label{eq:rho_fast}
  \rho_{\mathrm{f}} =
  \frac{\mathrm{E}[T_{\mathrm{f}}]}{\mathrm{E}[T_{\mathrm{cycle}}]} =
  \frac{\mathrm{E}[T_{\mathrm{f}}]}{\mathrm{E}[T_{\mathrm{f}}] +
    \mathrm{E}[T_{\mathrm{d}}] + \mathrm{E}[T_{\mathrm{tr}}] +
    \mathrm{E}[T_{\mathrm{on}}]}.
\end{equation}
In addition, we know that the utilization factor~$\rho$ satisfies
$\rho = \mathrm{E}[T_{\mathrm{on}}]/\mathrm{E}[T_{\mathrm{cycle}}] =
\mathrm{E}[T_{\mathrm{on}}] / (\mathrm{E}[T_{\mathrm{f}}] +
\mathrm{E}[T_{\mathrm{d}}] + \mathrm{E}[T_{\mathrm{tr}}] +
\mathrm{E}[T_{\mathrm{on}}])$ since, as long as the interface is
awake, it is transmitting queued frames. Rearranging terms,
$\mathrm{E}[T_{\mathrm{on}}]=\rho (\mathrm{E}[T_{\mathrm{f}}] +
\mathrm{E}[T_{\mathrm{d}}] + \mathrm{E}[T_{\mathrm{tr}}]) / (1-\rho)$
and, substituting this into~\eqref{eq:rho_fast}, we get
\begin{equation}
  \label{eq:rho_fast_2}
  \rho_{\mathrm{f}} = (1-\rho)
  \frac{E[T_{\mathrm{f}}]}{\mathrm{E}[T_{\mathrm{f}}] +
    \mathrm{E}[T_{\mathrm{d}}] + \mathrm{E}[T_{\mathrm{tr}}]}.
\end{equation}
Following the same approach, $\rho_{\mathrm{d}}$ can be calculated as
\begin{equation}
  \label{eq:rho_deep_2}
  \rho_{\mathrm{d}} = (1-\rho)
  \frac{E[T_{\mathrm{d}}]}{\mathrm{E}[T_{\mathrm{f}}] +
    \mathrm{E}[T_{\mathrm{d}}] + \mathrm{E}[T_{\mathrm{tr}}]}.
\end{equation}
Finally, substituting \eqref{eq:rho_fast_2} and \eqref{eq:rho_deep_2}
into~\eqref{eq:varphi}, we obtain that
\begin{equation}
  \label{eq:varphi_2}
  \varphi = 1 - (1 - \rho) \frac{(1 -
    \varphi_{\mathrm{f}})\mathrm{E}[T_{\mathrm{f}}] + (1 -
    \varphi_{\mathrm{d}})\mathrm{E}[T_{\mathrm{d}}]}{\mathrm{E}[T_{\mathrm{f}}]
    + \mathrm{E}[T_{\mathrm{d}}] + \mathrm{E}[T_{\mathrm{tr}}]}.
\end{equation}

In short, to compute the energy consumption of a dual-mode
EEE~interface, the average lengths of the Fast-Wake, Deep-Sleep and
transitioning periods must be obtained.

\subsection{Average Duration of Transitioning Periods}
\label{sec:trans}

Every coalescing cycle starts with a transition from active to the
Fast-Wake mode of length $T_{\mathrm{AtoF}}$. Then, if the interface
never enters the Deep-Sleep mode during the cycle, it will eventually
resume normal operation directly from Fast-Wake after a transition of
length $T_{\mathrm{FtoA}}$. Conversely, if the Deep-Sleep mode is
reached after a transition from Fast-Wake ($T_{\mathrm{FtoD}}$), the
interface will finally return to active after a longer transition of
length $T_{\mathrm{DtoA}}$. Therefore, the average duration of the
transitioning periods can be obtained as
\begin{equation}
  \label{eq:T_trans}
  \mathrm{E}[T_{\mathrm{tr}}] = T_{\mathrm{AtoF}} + (T_{\mathrm{FtoD}} + T_{\mathrm{DtoA}}) p_{\mathrm{d}} + T_{\mathrm{FtoA}} (1-p_{\mathrm{d}}),
\end{equation}
where $p_{\mathrm{d}}$ is the probability that the interface enters
the Deep-Sleep mode. Recall that the interface transitions to
Deep-Sleep just after a $T_{\mathrm{idle}}$~period in the Fast-Wake
mode if less than $Q_{\mathrm{f}}$~frames are queued for transmission,
so
\begin{equation}
  \label{eq:p_deep}
  p_{\mathrm{d}} = \sum_{i=0}^{Q_{\mathrm{f}}-1} \mathrm{Pr}[A(T_{\mathrm{AtoF}} + T_{\mathrm{idle}}) = i],
\end{equation}
where $A(\tau)$ is the number of frame arrivals in a time interval of
duration~$\tau$. 

\subsection{Average Duration of Fast-Wake Periods}
\label{sec:fast}

The duration of Fast-Wake periods depends on the arrival time of the
$Q_{\mathrm{f}}$-th~frame. If this frame arrives before finishing the
transition to Fast-Wake, then the interface will return to the active
state without entering Fast-Wake at all. Conversely, the interface
will stay in Fast-Wake until this frame arrives or the
$T_{\mathrm{idle}}$~timer expires, whatever happens first. Therefore,
the mean duration of Fast-Wake periods is given by
\begin{equation}
    \label{eq:T_fast}
  \mathrm{E}[T_{\mathrm{f}}] =
  \int_{T_{\mathrm{AtoF}}}^{T_{\mathrm{AtoF}}+T_{\mathrm{idle}}} \! (t - T_{\mathrm{AtoF}}) f_{Q_{\mathrm{f}}}(t) \, \mathrm{d}t 
  + \int_{T_{\mathrm{AtoF}}+T_{\mathrm{idle}}}^{\infty} \! T_{\mathrm{idle}} f_{Q_{\mathrm{f}}}(t) \, \mathrm{d}t ,
\end{equation}
where $f_{Q_{\mathrm{f}}}(t)$ is the probability density function
(pdf) of the time elapsed since the beginning of the coalescing cycle
until the arrival of the $Q_{\mathrm{f}}$-th~frame. Since we are
assuming independent and identically distributed interarrival times,
$\{I_n\}$ is a renewal process and $f_{Q_{\mathrm{f}}}(t) =
f_{T_{\mathrm{e}}}(t) \ast f_{I_2}(t) \ast \ldots \ast f_{I_{Q_f}}(t)
= f_{T_{\mathrm{e}}}(t) \ast f_I(t)^{\ast (Q_f-1)}$, where $\ast$ is
the convolution operator, $f_{T_{\mathrm{e}}}(t)$ is the pdf of the
empty periods and $f_I(t)$ is the pdf of the time between the arrivals
of two consecutive frames.

\subsection{Average Duration of Deep-Sleep Periods}
\label{sec:deep}

An interface in the Fast-Wake mode will only transition to Deep-Sleep
when it receives less than $Q_{\mathrm{f}}$~frames during the
$T_{\mathrm{AtoF}} + T_{\mathrm{idle}}$ period. Then, assuming that
just $i$~frames, $i<Q_{\mathrm{f}}$, are received during the Fast-Wake
period, the interface will remain in the Deep-Sleep mode until
$Q_{\mathrm{d}}-i$ more frames arrive. Therefore, with independent
interarrival times, the mean duration of Deep-Sleep periods can be
calculated as
\begin{equation}
    \label{eq:T_deep}
  \mathrm{E}[T_{\mathrm{d}}] = \sum_{i=0}^{Q_{\mathrm{f}}-1} \mathrm{Pr}[A(T_{\mathrm{AtoF}} + T_{\mathrm{idle}}) = i] \int_{T_{\mathrm{FtoD}}}^{\infty} \! (t - T_{\mathrm{FtoD}}) f_{Q_{\mathrm{d}}-i}(t) \, \mathrm{d}t,
\end{equation}
where $f_{Q_{\mathrm{d}}-i}(t) = f_{I_{i+1}}(t) \ast f_{I_{i+2}}(t)
\ast \ldots \ast f_{I_{Q_d}}(t) = f_I(t)^{\ast (Q_d-i)}$ is the pdf of
the time elapsed since the arrival of the $i$-th~frame until the
$Q_{\mathrm{d}}$-th~frame arrives.

\section{Poisson Traffic}
\label{sec:poisson_model}

In this section we will calculate the average lengths of the
Fast-Wake, Deep-Sleep and transitioning periods when frame arrivals
follow a Poisson process. Although it is well-known that frame
arrivals on LAN networks do not really follow a Poisson
distribution~\cite{leland94:on_the_self_similar}, Poisson traffic is
useful in order to provide a valid approximation to aggregated traffic
in the Internet core~\cite{vishwanath09:_how_poisson_is_tcp_traffic}.

According to Sect.~\ref{sec:trans}, to obtain the average length of
the transitioning periods, it only remains to calculate the
probability that the interface enters the Deep-Sleep mode
($p_{\mathrm{d}}$). It is well known that, with Poisson traffic,
$\mathrm{Pr}[A(\tau)=i] = \mathrm{e}^{-\lambda \tau} (\lambda \tau)^i
/ i!$, so, substituting this into~\eqref{eq:p_deep}, we get
\begin{equation}
  \label{eq:p_deep_poisson}
  p_{\mathrm{d}} = R(Q_{\mathrm{f}}, \lambda(T_{\mathrm{AtoF}} + T_{\mathrm{idle}})) 
  = \frac{\Gamma (Q_{\mathrm{f}}, \lambda(T_{\mathrm{AtoF}} + T_{\mathrm{idle}}))}{\Gamma (Q_{\mathrm{f}})},
\end{equation}
where $R(q,x)$ is the regularized upper incomplete gamma function,
$\Gamma (q,x) = \int_x^{\infty} t^{q-1} \mathrm{e}^{-t}\, \mathrm{d}t$
is the upper incomplete gamma function and $\Gamma (q) = \Gamma
(q,0)$.

Regarding the average duration of Fast-Wake periods, note that, due to
the memoryless property of Poisson traffic, $f_{T_{\mathrm{e}}}(t) =
f_I(t)$, so $f_{Q_{\mathrm{f}}}(t) = f_I(t)^{\ast Q_f}$. Additionally,
since for Poisson traffic all interarrival times are exponentially
and identically distributed, the arrival time of the
$Q_{\mathrm{f}}$-th frame is Erlang-$Q_{\mathrm{f}}$ distributed and
$f_{Q_{\mathrm{f}}}(t) = \lambda^{Q_{\mathrm{f}}} t^{Q_{\mathrm{f}}-1}
\mathrm{e}^{-\lambda t} / (Q_{\mathrm{f}}-1)!$ Using this to
solve~\eqref{eq:T_fast}, we get
\begin{IEEEeqnarray}{rCl}
  \label{eq:T_fast_poisson}
  \mathrm{E}[T_{\mathrm{f}}] &=& Q_{\mathrm{f}}\,\bigl( R(Q_{\mathrm{f}}+1, \lambda T_{\mathrm{AtoF}})-R(Q_{\mathrm{f}}+1, \lambda(T_{\mathrm{AtoF}} + T_{\mathrm{idle}})\bigr)/\lambda \nonumber \\
  & & - \, T_{\mathrm{AtoF}}\,R(Q_{\mathrm{f}}, \lambda T_{\mathrm{AtoF}}) + (T_{\mathrm{AtoF}}+T_{\mathrm{idle}}) p_{\mathrm{d}}.
\end{IEEEeqnarray}

Finally, applying $f_{Q_{\mathrm{d}}-i}(t) =
\lambda^{Q_{\mathrm{d}}-i} t^{Q_{\mathrm{d}}-i-1} \mathrm{e}^{-\lambda
  t} / (Q_{\mathrm{d}}-i-1)!$ into~\eqref{eq:T_deep}, we obtain that
the average duration of Deep-Sleep periods is given by
\begin{flalign}
    \label{eq:T_deep_poisson}
  & \mathrm{E}[T_{\mathrm{d}}] = \sum_{i=0}^{Q_{\mathrm{f}}-1} \frac{\mathrm{e}^{-\lambda(T_{\mathrm{AtoF}} + T_{\mathrm{idle}})} (\lambda(T_{\mathrm{AtoF}} + T_{\mathrm{idle}}))^i}{i!}& \nonumber \\
    & \cdot \bigl( (Q_{\mathrm{d}}-i)\,R(Q_{\mathrm{d}}-i+1, \lambda T_{\mathrm{FtoD}})/\lambda - T_{\mathrm{FtoD}}\,R(Q_{\mathrm{d}}-i, \lambda T_{\mathrm{FtoD}}) \bigr).&
\end{flalign}

\section{Evaluation}
\label{sec:evaluation}

To evaluate the performance of the proposed scheme and validate our
model, we conducted several experiments on an in-house simulator,
available for download at~\cite{herreria15:dualmode_eee_simulator}. We
simulated a $40\,$Gb$/$s interface receiving Poisson traffic with an
average arrival rate varying from $2$ to $38\,$Gb$/$s. The frame size
was set to $1500\,$bytes. Regarding the PHY features, we set the
transition times and the efficiency profile to the same values as
in~\cite{mostowfi15:dual_mode_eee}: $T_{\mathrm{AtoF}}=0.90\,\mu$s,
$T_{\mathrm{FtoA}}=0.34\,\mu$s, $T_{\mathrm{FtoD}}=1.00\,\mu$s,
$T_{\mathrm{DtoA}}=5.50\,\mu$s, $T_{\mathrm{idle}}=3.50\,\mu$s,
$\varphi_{\mathrm{f}}=0.7$ and $\varphi_{\mathrm{d}}=0.1$.

\begin{figure}[t]
  \centering
  \subfigure[Energy consumption.]{
    \includegraphics[width=0.9\columnwidth]{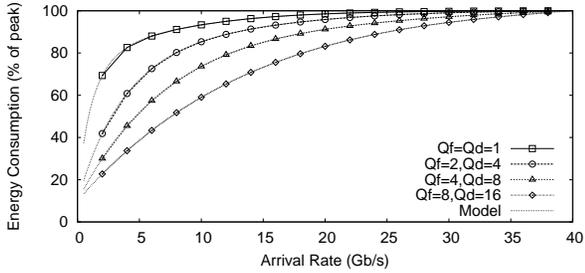}
    \label{fig:poisson-energy}
  }
  \subfigure[Average queueing delay.]{
    \includegraphics[width=0.9\columnwidth]{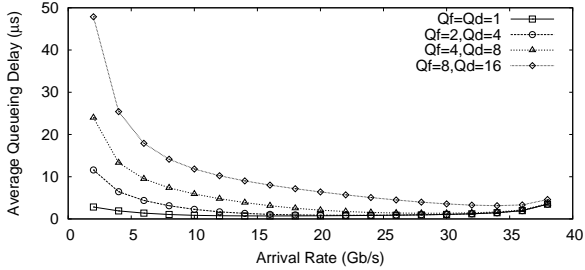}
    \label{fig:poisson-delay}
  }
  \caption{Results with Poisson traffic.}
  \label{fig:poisson}
\end{figure}

Figure~\ref{fig:poisson} shows both the energy consumption and the
average queueing delay obtained without coalescing ($Q_{\mathrm{f}} =
Q_{\mathrm{d}} = 1$) and when coalescing is applied with three
different queue threshold configurations.\footnote{Each simulation was
  run for ten seconds and repeated ten times using different random
  seeds. The average and $95\,\%$ confidence intervals (CIs) of every
  performance measure were calculated but CIs have not been
  represented in the graphs since all of them are small enough and
  just clutter the figures.}  Note that our model provides very
accurate predictions for the energy consumption in all the simulated
scenarios. As expected, frame coalescing significantly reduces energy
consumption at the expense of increasing frame latency and the higher
the queue thresholds are, the greater energy savings and frame delays
are obtained.

\begin{figure}[t]
  \centering 
  \subfigure[Energy consumption. Filled points show model predictions.]{
    \includegraphics[width=0.9\columnwidth]{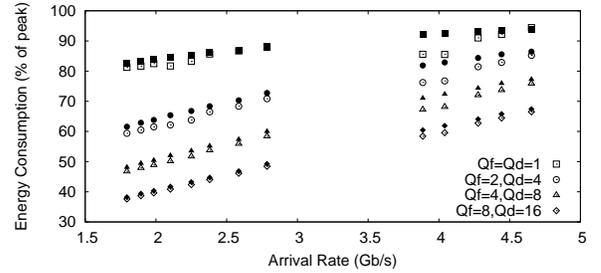}
    \label{fig:caida-energy}
  }
  \subfigure[Average queueing delay.]{
    \includegraphics[width=0.9\columnwidth]{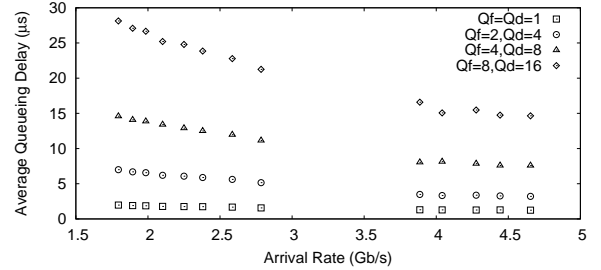}
    \label{fig:caida-delay}
  }
  \caption{Results for CAIDA traces.}
  \label{fig:caida}
\end{figure}

We also evaluated frame coalescing using real world traffic traces
publicly available from the CAIDA archive~\cite{caida15}. The analyzed
CAIDA traces were collected during 2015 on a $10\,$Gb/s backbone
Ethernet link. Though $10$~Gigabit links only have a single LPI mode,
we assumed that the traced PHY behaves as the previously simulated
dual-mode interface and uses the same configuration
settings. Figure~\ref{fig:caida} shows the obtained results. Again,
frame coalescing allows important reductions on energy consumption at
the cost of tolerable increments on frame
delay. Fig.~\ref{fig:caida-energy} also shows the energy consumption
predicted by our Poisson model. Not surprisingly, the model
predictions are quite similar to the measured values, thus confirming
the Poissonian-like nature of aggregated traffic.

\section{Conclusions}
\label{sec:conclusions}

This paper provides an accurate model for dual-mode EEE interfaces
that can be used to analyze the influence of frame coalescing
parameters and PHY characteristics on their energy
consumption. Simulation results for both synthetic and real Internet
traffic traces assess the validity of our model and confirm that frame
coalescing allows significant energy savings at the cost of some
tolerable increments on frame delay.

\section*{Acknowledgments}

Support for CAIDA's Internet Traces is provided by the National
Science Foundation, the US Department of Homeland Security, and CAIDA
Members.  


\bibliographystyle{IEEEtran}
\bibliography{IEEEabrv,dual_mode_eee}

\begin{thebibliography}{10}
\providecommand{\url}[1]{#1}
\csname url@samestyle\endcsname
\providecommand{\newblock}{\relax}
\providecommand{\bibinfo}[2]{#2}
\providecommand{\BIBentrySTDinterwordspacing}{\spaceskip=0pt\relax}
\providecommand{\BIBentryALTinterwordstretchfactor}{4}
\providecommand{\BIBentryALTinterwordspacing}{\spaceskip=\fontdimen2\font plus
\BIBentryALTinterwordstretchfactor\fontdimen3\font minus
  \fontdimen4\font\relax}
\providecommand{\BIBforeignlanguage}[2]{{%
\expandafter\ifx\csname l@#1\endcsname\relax
\typeout{** WARNING: IEEEtran.bst: No hyphenation pattern has been}%
\typeout{** loaded for the language `#1'. Using the pattern for}%
\typeout{** the default language instead.}%
\else
\language=\csname l@#1\endcsname
\fi
#2}}
\providecommand{\BIBdecl}{\relax}
\BIBdecl

\bibitem{802.3az}
\BIBentryALTinterwordspacing
``{IEEE} {S}td 802.3az-2010,'' Oct. 2010. [Online]. Available:
  \url{http://dx.doi.org/10.1109/IEEESTD.2010.5621025}
\BIBentrySTDinterwordspacing

\bibitem{christensen10:_the_road_to_eee}
K.~Christensen, P.~Reviriego, B.~Nordman, M.~Bennett, M.~Mostowfi, and J.~A.
  Maestro, ``{IEEE} 802.3az: the road to energy efficient {E}thernet,''
  \emph{{IEEE} Commun. Mag.}, vol.~48, no.~11, pp. 50--56, 2010.

\bibitem{herreria12:gig1}
S.~Herrería-Alonso, M.~Rodríguez-Pérez, M.~Fernández-Veiga, and
  C.~López-García, ``A {GI/G/1} model for 10 {Gb/s} energy efficient
  {Ethernet} links,'' \emph{{IEEE} Trans. Commun.}, vol.~60, no.~11, pp.
  3386--3395, Nov. 2012.

\bibitem{802.3bj}
\BIBentryALTinterwordspacing
``{IEEE} {S}td 802.3bj-2014 amendment 2: Physical layer specifications and
  management parameters for 100 {Gb/s} operation over backplanes and copper
  cables,'' Sep. 2014. [Online]. Available:
  \url{http://dx.doi.org/10.1109/IEEESTD.2014.6891095}
\BIBentrySTDinterwordspacing

\bibitem{barrass12:_options_100g}
H.~Barrass, ``Options for {EEE} in {100G},'' Presentation at {IEEE P802.3bj}
  meeting, Jan. 2012.

\bibitem{mostowfi15:dual_mode_eee}
M.~Mostowfi, ``A simulation study of energy efficient ethernet with two modes
  of low-power operation,'' \emph{{IEEE} Commun. Lett.}, Jul. 2015.

\bibitem{leland94:on_the_self_similar}
W.~E. Leland, M.~S. Taqqu, W.~Willinger, and D.~V. Wilson, ``On the
  self-similar nature of {E}thernet traffic (extended version),''
  \emph{{IEEE/ACM T}rans. {N}etw.}, vol.~2, no.~1, pp. 1--15, Feb. 1994.

\bibitem{vishwanath09:_how_poisson_is_tcp_traffic}
A.~Vishwanath, V.~Sivaraman, and D.~Ostry, ``How {Poisson} is {TCP} traffic at
  short time-scales in a small buffer core network?'' in \emph{Advanced
  Networks and Telecommunication Systems (ANTS), IEEE 3rd International
  Symposium on}, Dec. 2009.

\bibitem{herreria15:dualmode_eee_simulator}
\BIBentryALTinterwordspacing
S.~Herrería-Alonso, ``Dualmodeeeesimulator: a java program that simulates a
  dual-mode {EEE} link.'' [Online]. Available:
  \url{https://github.com/sherreria/DualModeEeeSimulator}
\BIBentrySTDinterwordspacing

\bibitem{caida15}
\BIBentryALTinterwordspacing
``The {CAIDA UCSD} anonymized {Internet} traces 2015 - {Dates} used: 20150219,
  20150521.'' [Online]. Available:
  \url{http://www.caida.org/data/passive/passive\_2015\_dataset.xml}
\BIBentrySTDinterwordspacing

\end{thebibliography}

\end{document}